\documentclass[journal]{IEEEtran}
\usepackage{cite,amsmath}
\usepackage{subfig}
\usepackage{color}
\usepackage{graphicx}
\usepackage{pstool}
\usepackage{array,booktabs}

\newcommand{\ve}[1]{\boldsymbol{#1}}
\newcommand{\te}[1]{\overline{\overline{#1}}}

\newcolumntype{N}{@{}m{0pt}@{}}

\newcounter{tempEquationCounter}
\newcounter{thisEquationNumber}
\newenvironment{floatEq}
{\setcounter{thisEquationNumber}{\value{equation}}\addtocounter{equation}{1}
\begin{figure*}[!t]
\normalsize\setcounter{tempEquationCounter}{\value{equation}}
\setcounter{equation}{\value{thisEquationNumber}}
}
{\setcounter{equation}{\value{tempEquationCounter}}
\hrulefill\vspace*{4pt}
\end{figure*}
}

\begin{document}

\title{Surface-Wave Dispersion Retrieval Method and Synthesis Technique for Bianisotropic Metasurfaces}

\author{Karim~Achouri, and Olivier J. F. Martin

\thanks{K. Achouri,  and O. J. F. Martin are with the Nanophotonics and Metrology Laboratory, Department of Microengineering, $\acute{\mathrm{E}}$cole Polytechnique F$\acute{\mathrm{e}}$d$\acute{\mathrm{e}}$rale de Lausanne, Route Cantonale, 1015 Lausanne, Switzerland (e-mail: karim.achouri@epfl.ch).}
}

\maketitle

\begin{abstract}
We propose a surface-wave dispersion retrieval method and synthesis technique that applies to bianisotropic metasurfaces that are embedded in symmetric or asymmetric environments. Specifically, we use general zero-thickness sheet transition conditions to relate the propagation constants of surface-wave modes to the bianisotropic susceptibility components of the metasurface, which can themselves be directly related to its scattering parameters. It is then possible to either obtain the metasurface dispersion diagram from its known susceptibilities or, alternatively, compute the susceptibilities required to achieve a desired surface-wave propagation. The validity of the method is demonstrated by comparing its results to those obtained with exact dispersion relations of well known structures such as the propagation of surface plasmons on thin metallic films.
In particular, this work reveals that it is possible to achieve surface-wave propagation only on one side of the metasurface either by superposition of symmetric and asymmetric modes in the case of anisotropic metasurfaces or by completely forbidding the existence of the surface wave on one side of the structure using bianisotropic metasurfaces.
\end{abstract}


\IEEEpeerreviewmaketitle


\section{Introduction}

Metasurfaces are thin arrays of artificial scattering particles engineered to control the propagation of light in ways that are unachievable with conventional matters~\cite{yu2014flat,capasso1,Pfeiffer2013Huygens,Kildishev1232009,pfeiffer2014high,achouri2014general,epstein2016arbitrary,asadchy2015functional,Achouri2015c,achouri2017design}. In recent years, several metasurface synthesis techniques have been developed to design these complex structures and allow one to exploit them to their full potential~\cite{pfeiffer2014high,achouri2014general,6477089,epstein2016arbitrary}. The main applications of these synthesis techniques have been to implement metasurfaces controlling electromagnetic waves in the far-field regime. In parallel, other synthesis techniques have been developed for the specific near-field problem of surface-wave propagation on artificial structures such as tensor impedance surfaces~\cite{1236096,6484911,6308707,elek2015synthesis,SurfaceWaveguide}. However, these techniques have been mostly limited to the case of impenetrable (opaque) surfaces and have also been restricted to the case of anisotropic material parameters. Overcoming, these two limitations is the initial motivation of this work.

Concretely, this paper aims at providing the reader with a general dispersion retrieval method for penetrable and impenetrable bianisotropic metasurfaces embedded in symmetric or asymmetric environments. This work is an extension of our metasurface synthesis technique developed in~\cite{achouri2014general} with the specific objective of studying the propagation of surface-wave modes. It allows one to compute the dispersion diagram of metasurfaces based on the knowledge of their effective surface susceptibilities, which can themselves be directly related to the metasurface scattering parameters. Our developments are based on the generalized sheet transition conditions, which represent the basis for our metasurface synthesis and analysis framework~\cite{achouri2017design,vahabzadeh2017computational}. Importantly, this work may also be used for the synthesis of surface-wave guiding metasurfaces since we also provide the analytical expressions of the dispersion relations in the case of birefringent and bianisotropic (omega-type) metasurfaces. This means that one may find the required effective susceptibilities so as to achieve a specified propagation constant and polarization of the surface-wave mode.

In order to evaluate the validity of the proposed dispersion retrieval method, we will use it to compute the dispersion diagram of several well known structures for which there exist exact expressions of their dispersion relations. This will allow us to compare the dispersion diagram predicted by the proposed method to that corresponding to these exact solutions. Specifically, we will consider the propagation of surface plasmon polaritons on (a)symmetric thin metallic layers, which have been already widely studied~\cite{SaridLongRange,raether1988surface,Berini:09,Akimov2017}.

This paper is organized as follows: Sec.~\ref{sec:model} introduces the modelling of bianisotropic metasurfaces through sheet transition conditions and presents the formulation of the surface-wave dispersion retrieval method. Sec.~\ref{sec:sym} applies this method to the situation of symmetric environments, i.e. when the metasurface is surround by the same medium on both sides. This section illustrates the application of the dispersion retrieval method with two examples: birefringent and bianisotropic metasurfaces. Then, Sec.~\ref{sec:asym} discusses the case of asymmetric environments. Note that the proposed dispersion retrieval method is based on a a priori knowledge of the metasurface effective susceptibilities. For the sake of completeness, Appendix~\ref{app:Sparam} provides a summarized discussion of the susceptibility retrieval method presented in~\cite{achouri2017design} and which allows one to compute the effective susceptibilities of a metasurface from its scattering parameters.

\section{Modelling of Bianisotropic Metasurfaces}
\label{sec:model}

Metasurfaces may be conveniently and effectively modelled by zero-thickness sheets of polarizable electric and magnetic dipolar moments~\cite{kuester2003av,pfeiffer2014high,achouri2014general,asadchy2015functional,epstein2016arbitrary,michele2017}. The most general transition conditions that apply to this type of electromagnetic discontinuity were initially derived by Idemen~\cite{Idemen1973}. They have been later successfully applied to the case of metasurfaces and have been since then referred to as the generalized sheet transition conditions (GSTCs)~\cite{kuester2003av,achouri2014general}. For a metasurface lying in the $xy$-plane at $z=0$, the GSTCs with time dependence $e^{j\omega t}$ read
\begin{subequations}
\label{eq:GSTCs}
\begin{equation}
\hat{\ve{z}}\times\Delta\ve{H}=\frac{\partial}{\partial t}\ve{P} - \hat{\ve{z}}\times \nabla M_z,
\end{equation}
\begin{equation}
\hat{\ve{z}}\times\Delta\ve{E}=-\mu\frac{\partial}{\partial t}\ve{M} - \frac{1}{\epsilon}\hat{\ve{z}}\times \nabla P_z,
\end{equation}
\end{subequations}
where $\Delta\ve{E}$ and $\Delta\ve{H}$ are the differences of the electric and magnetic fields on both sides of the metasurface, and $\ve{P}$ and $\ve{M}$ refer to the electric and magnetic polarization densities, respectively. Note that, throughout the paper, the material parameters, such as the permittivity, as well as the wavenumbers that do not present any numeral subscript refer to vacuum.

While relations~\eqref{eq:GSTCs} are general, they are not necessarily practical to use in their current form due to the presence of the spatial derivatives of the normal components of the polarization densities. Consequently, it has been common practice to ignore the presence of normal polarizations within metasurfaces and thus assume that $P_z = M_z = 0$ so as to transform relations~\eqref{eq:GSTCs} into algebraic equations. This simplification is, in most cases, justified by the fact that metasurfaces are so thin compared to the operating wavelength that normal polarizations are negligible compared to their tangential counterparts. However, in some particular cases, the assumption that $P_z = M_z = 0$ is not valid (e.g. metallic loops in the $xy$-plane generating strong $M_z$ components), which produces discrepancies between the purely tangential GSTCs model and the actual electromagnetic responses of metasurfaces. In this work, we will nonetheless use the assumption that $P_z = M_z = 0$ for the sake of simplicity and will discuss the implications of this simplification with several examples. Note that it is possible to solve the general relations~\eqref{eq:GSTCs} in order to compute the dispersion relations of metasurfaces without the assumption of purely tangential polarizations. However, the additional complexity that this would entail is beyond the scope of this paper and may be the topic of future work.

We now remove the spatial derivatives of the normal polarizations and express the tangential components of $\ve{P}$ and $\ve{M}$ in terms of the metasurface bianisotropic surface susceptibilities such that~\eqref{eq:GSTCs} transforms into~\cite{achouri2014general}
\begin{subequations}
\label{eq:InvProb}
\begin{align}
\ve{\hat{z}}\times\Delta\ve{H}
&=j\omega\epsilon\te{\chi}_\text{ee}\cdot\ve{E}_\text{av}+jk\te{\chi}_\text{em}\cdot\ve{H}_\text{av},\label{eq:diffH}\\
\ve{\hat{z}}\times\Delta\ve{E}
&=-j\omega\mu \te{\chi}_\text{mm}\cdot\ve{H}_\text{av}-jk\te{\chi}_\text{me}\cdot\ve{E}_\text{av},\label{eq:diffE}
\end{align}
\end{subequations}
where $\ve{E}_\text{av}$ and $\ve{H}_\text{av}$ are the arithmetic average of the electric and magnetic fields across the metasurface, respectively.

The objective of this work is to obtain the dispersion relations of metasurfaces. To achieve this goal, we will solve~\eqref{eq:InvProb} so as to find the surface-wave modes, supported by the metasurface, as a function of its susceptibilities. In this problem, the susceptibilities are considered to be known quantities, while the propagation constants and the polarizations of the surface waves are unknown. Therefore, the first step in order to obtain dispersion relations of a metasurface is to find its effective susceptibilities, which can be easily done using conventional homogenization techniques~\cite{pfeiffer2014high,achouri2014general,asadchy2015functional,epstein2016arbitrary,achouri2017design}. For the completeness of this work, the susceptibility retrieval technique presented in~\cite{achouri2017design} is summarized in Appendix~\ref{app:Sparam}.
\begin{figure}[h!]
\centering
\includegraphics[width=0.9\columnwidth]{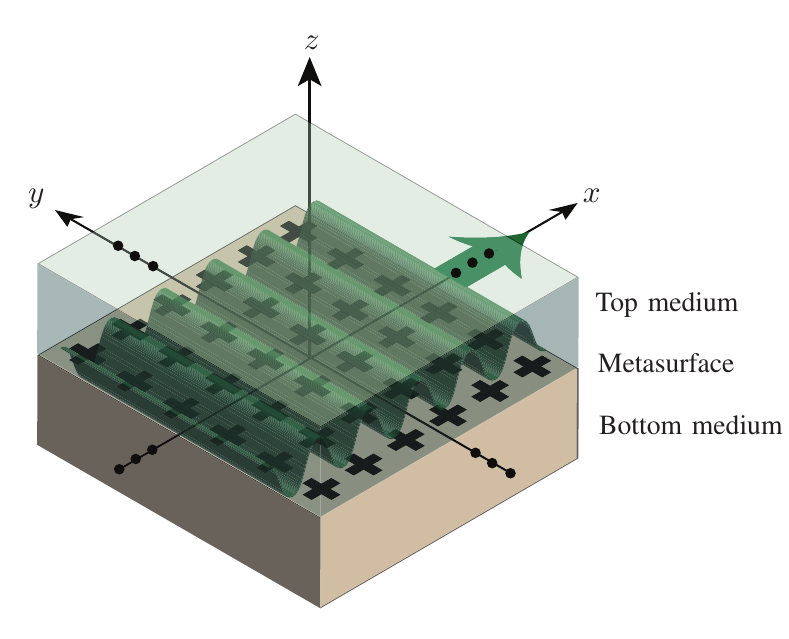}
\caption{Surface wave propagating on a metasurface surrounded by two different media.}
\label{fig:meta}
\end{figure}

Now that we have a way to find the metasurface susceptibilities, we transform~\eqref{eq:InvProb} to be able to extract the polarizations and propagation constants of the surface-wave modes. To do so, we specify the electromagnetic fields corresponding to surface waves propagating on the top and on the bottom of the metasurface. Note that in the forthcoming developments, we will only consider the propagation of surface waves in the $x$-direction, as depicted in Fig.~\ref{fig:meta}. Nevertheless, the propagation of surface waves in any direction in the $xy$-plane may be easily obtained by simple matrix rotation of the susceptibility tensors. For instance, each of the susceptibility tensors in~\eqref{eq:InvProb} may be rotated using $\te{\chi}_\text{rot} = \te{R}\cdot\te{\chi}\cdot\te{R}^{-1}$, where $\te{R}$ is the rotation matrix~\cite{arfken1999mathematical}.

In a very general situation, the electromagnetic fields on the top ($z=0^+$) and on the bottom ($z=0^-$) of the metasurface may be expressed as superpositions of transverse electric (TE) and transverse magnetic (TM) modes. In addition, we consider here the case of a source-less (or undriven) problem, which means that we do not have to specify any excitation. Accordingly, the fields on the $z=0^-$ side of the metasurface are given by
\begin{subequations}\label{eq:BottomFields}
\begin{align}
\ve{E}^{0^-} &= \left(-\hat{\ve{x}}\frac{k_{1z}}{k_1}\eta_1 A_\text{TM}^{0^-} + \hat{\ve{y}}A_\text{TE}^{0^-}\right)e^{-jk_x x}, \\
\ve{H}^{0^-} &= \left(\hat{\ve{x}}\frac{k_{1z}}{k_1\eta_1} A_\text{TE}^{0^-} + \hat{\ve{y}}A_\text{TM}^{0^-}\right)e^{-jk_x x},
\end{align}
\end{subequations}
where $k_1^2 = k_{1z}^2 + k_x^2$ and $A_\text{TE}^{0^-}$ and $A_\text{TM}^{0^-}$ are the complex amplitudes of the TE and TM surface-wave modes on the bottom of the metasurface. Similarly, the fields on the $z=0^+$ side of the metasurface are given by
\begin{subequations}\label{eq:TopFields}
\begin{align}
\ve{E}^{0^+} &= \left(\hat{\ve{x}}\frac{k_{2z}}{k_2}\eta_2 A_\text{TM}^{0^+} + \hat{\ve{y}}A_\text{TE}^{0^+}\right)e^{-jk_x x}, \\
\ve{H}^{0^+} &= \left(-\hat{\ve{x}}\frac{k_{2z}}{k_2\eta_2} A_\text{TE}^{0^+} + \hat{\ve{y}}A_\text{TM}^{0^+}\right)e^{-jk_x x},
\end{align}
\end{subequations}
where $k_2^2 = k_{2z}^2 + k_x^2$ and $A_\text{TE}^{0^+}$ and $A_\text{TM}^{0^+}$ are the complex amplitudes of the TE and TM surface-wave modes on the top of the metasurface. Note that the propagation constant $k_x$ is, by phase matching, the same for the surface waves on the top and the bottom of the metasurface.

The difference and the average of the fields in~\eqref{eq:InvProb} may now be expressed in terms of the electromagnetic fields given in~\eqref{eq:BottomFields} and~\eqref{eq:TopFields}. They respectively read
\begin{subequations}\label{eq:Diff}
\begin{align}
\Delta\ve{E} &= \ve{E}^{0^+} - \ve{E}^{0^-},\\
\Delta\ve{H} &= \ve{H}^{0^+} - \ve{H}^{0^-},
\end{align}
\end{subequations}
and
\begin{subequations}\label{eq:Av}
\begin{align}
\ve{E}_\text{av} &= \frac{1}{2}\left(\ve{E}^{0^+} + \ve{E}^{0^-} \right),\\
\ve{H}_\text{av} &= \frac{1}{2}\left(\ve{H}^{0^+} + \ve{H}^{0^-}\right).
\end{align}
\end{subequations}
Finally, we transform the GSTCs relations~\eqref{eq:InvProb} by substituting~\eqref{eq:Diff} and~\eqref{eq:Av} into~\eqref{eq:InvProb} along with~\eqref{eq:BottomFields} and~\eqref{eq:TopFields}. After rearranging the terms and simplifying the expressions, we obtain the following system of equation:
\begin{equation}
\label{eq:GeneralEq}
\te{\chi}\cdot\ve{x} = 0,
\end{equation}
where the matrix $\te{\chi}$ is defined in Eq.~\eqref{eq:ChiMat} below and the vector $\ve{x}$ reads
\begin{equation}\label{eq:x}
\ve{x}^\text{T} = (A_\text{TE}^{0^-}, A_\text{TM}^{0^-},A_\text{TE}^{0^+}, A_\text{TM}^{0^+}),
\end{equation}
with ``T'' being the transpose operation.

In what follows, we will show how the general system~\eqref{eq:GeneralEq} may be solved to obtain the dispersion relations of different types of structures.

\begin{floatEq}
\begin{equation}\label{eq:ChiMat}
\te{\chi}=
\begin{pmatrix}
  j\omega\epsilon\chi_\text{ee}^{xy}+\frac{jk_{1z}k\chi_\text{em}^{xx}}{k_1 \eta_1} & -2+jk\chi_\text{em}^{xy}-\frac{jk_{1z}\eta_1\omega\epsilon\chi_\text{ee}^{xx}}{k_1} & j\omega\epsilon\chi_\text{ee}^{xy}-\frac{jk_{2z}k\chi_\text{em}^{xx}}{k_2\eta_2} & 2+jk\chi_\text{em}^{xy} +\frac{jk_{2z}\eta_2\omega\epsilon\chi_\text{ee}^{xx}}{k_2} \\
  j\omega\epsilon\chi_\text{ee}^{yy}+\frac{k_{1z}(2+jk\chi_\text{em}^{yx})}{k_1\eta_1} & jk\chi_\text{em}^{yy}-\frac{jk_{1z}\eta_1^2\omega\epsilon\chi_\text{ee}^{yx}}{k_1\eta_1} & j\omega\epsilon\chi_\text{ee}^{yy}+\frac{k_{2z}(2-jk\chi_\text{em}^{yx})}{k_2\eta_2} & jk\chi_\text{em}^{yy}+\frac{jk_{2z}\eta_2^2\omega\epsilon\chi_\text{ee}^{yx}}{k_2\eta_2} \\
  -2-jk\chi_\text{me}^{xy}-\frac{jk_{1z}\omega\mu\chi_\text{mm}^{xx}}{k_1 \eta_1} & -j\omega\mu\chi_\text{mm}^{xy}+\frac{jk_{1z}\eta_1k\chi_\text{me}^{xx}}{k_1} & 2-jk\chi_\text{me}^{xy}+\frac{jk_{2z}\omega\mu\chi_\text{mm}^{xx}}{k_2 \eta_2} & -j\omega\mu\chi_\text{mm}^{xy}-\frac{jk_{2z}\eta_2k\chi_\text{me}^{xx}}{k_2} \\
  -jk\chi_\text{me}^{yy}-\frac{jk_{1z}\omega\mu\chi_\text{mm}^{yx}}{k_1 \eta_1} & -j\omega\mu\chi_\text{mm}^{yy}-\frac{k_{1z}\eta_1(2-jk\chi_\text{me}^{yx})}{k_1} & -jk\chi_\text{me}^{yy}+\frac{jk_{2z}\omega\mu\chi_\text{mm}^{yx}}{k_2 \eta_2} & -j\omega\mu\chi_\text{mm}^{yy} -\frac{k_{2z}\eta_2(2+jk\chi_\text{me}^{yx})}{k_2}
\end{pmatrix}.
\end{equation}
\end{floatEq}

As a side note, we provide here the Lorentz reciprocity conditions, which will be of practical interest for the forthcoming developments since they relate the susceptibility tensors in the following fashion~\cite{rothwell2008electromagnetics}:
\begin{equation}
\label{eq:reciprocity}
\te{\chi}_\text{ee}^\text{T}=\te{\chi}_\text{ee},\qquad
\te{\chi}_\text{mm}^\text{T}=\te{\chi}_\text{mm},\qquad
\te{\chi}_\text{me}^\text{T}=-\te{\chi}_\text{em}.
\end{equation}
A direct consequence of these reciprocity conditions is that they reduce the number of degrees of freedom available to control the propagation of surface waves, by relating several susceptibility components to each other in~\eqref{eq:ChiMat}.

\section{Dispersion in a Symmetric Environment}
\label{sec:sym}

We start by discussing the case of symmetric environments, which means that the media on both sides of the metasurface are the same and thus that $\eta = \eta_1 = \eta_2$, $k = k_1 = k_2$, and $k_z = k_{1z} = k_{2z}$. This greatly simplifies the system~\eqref{eq:GeneralEq}, which may now be written in the form of an eigenvalue problem to be subsequently solved for the metasurface dispersion relations. Specifically, we split the matrix $\te{\chi}$ in~\eqref{eq:ChiMat} into a matrix $\te{A}$, which does not contain $k_z$, and a matrix $\te{B}$, which does. The system~\eqref{eq:GeneralEq} thus transforms into
\begin{equation}\label{eq:Eig}
\te{A}\cdot\ve{x} = k_z \te{B}\cdot\ve{x},
\end{equation}
where the matrices $\te{A}$ and $\te{B}$ are given in Eqs.~\eqref{eq:MatAB} below.
\begin{subequations}\label{eq:MatAB}
\begin{floatEq}
\begin{equation}\label{eq:MatA}
\te{A}=
\begin{pmatrix}
  -k \chi_\text{ee}^{xy} & -2j \eta -k \eta \chi_\text{em}^{xy} & -k \chi_\text{ee}^{xy} & 2j\eta - k \eta \chi_\text{em}^{xy} \\
  k^2 \chi_\text{ee}^{yy} & k^2 \eta \chi_\text{em}^{yy} & k^2 \chi_\text{ee}^{yy} & k^2\eta \chi_\text{em}^{yy} \\
  -2j+k\chi_\text{me}^{xy} & k\eta \chi_\text{mm}^{xy} & 2j + k\chi_\text{me}^{xy} & k\eta\chi_\text{mm}^{xy} \\
  k^2\chi_\text{me}^{yy} & k^2\eta\chi_\text{mm}^{yy} & k^2\chi_\text{me}^{yy} & k^2\eta\chi_\text{mm}^{yy}
\end{pmatrix},
\end{equation}
\begin{equation}\label{eq:MatB}
\te{B}=
\begin{pmatrix}
  \chi_\text{em}^{xx} & -\eta \chi_\text{ee}^{xx} & -\chi_\text{em}^{xx} & \eta \chi_\text{ee}^{xx} \\
  2j -k\chi_\text{em}^{yx} & k\eta\chi_\text{ee}^{yx} & 2j+k\chi_\text{em}^{yx} & -k\eta\chi_\text{ee}^{yx} \\
  -\chi_\text{mm}^{xx} & \eta \chi_\text{me}^{xx} & \chi_\text{mm}^{xx} & -\eta\chi_\text{me}^{xx} \\
  -k\chi_\text{mm}^{yx} & 2j\eta+k\eta\chi_\text{me}^{yx} & k\chi_\text{mm}^{yx} & 2j\eta-k\eta\chi_\text{me}^{yx}
\end{pmatrix}.
\end{equation}
\end{floatEq}
\end{subequations}
Then, by matrix inversion of $\te{B}$, we obtain the following eigenvalue problem
\begin{equation}\label{eq:Eig2}
\te{M}\cdot\ve{x} = k_z \ve{x},
\end{equation}
where the matrix $\te{M}$ is simply given by $\te{M}=\te{B}^{-1}\cdot\te{A}$. This equation may now be solved to obtain the eigenvalue $k_z$, which is directly related to the propagation constant using $k^2 = k_z^2 + k_x^2$, and the corresponding eigenvector $\ve{x}$, which provides the amplitude of the TE and TM surface-wave modes supported by the metasurface. It follows that, if the metasurface susceptibilities are known, it is trivial to retrieve the metasurface dispersion diagram.

In what follows, we will analytically solve the eigenvalue problem~\eqref{eq:Eig2} for the two particularly common and important cases of: (A) birefringence and (B) bianisotropy (omega-type).

\subsection{Dispersion of Birefringent Metasurfaces}

Birefringent metasurfaces are anisotropic structures, which represent the most common class of metasurfaces that have been studied and reported in the literature so far~\cite{Glybovski20161}. Accordingly, they are expected to be the most likely candidate for controlling the propagation of surface waves. They have the advantage of being relatively simple to model since they only possess the following nonzero susceptibilities components $\chi_\text{ee}^{xx},~\chi_\text{ee}^{yy},~\chi_\text{mm}^{xx}$ and $\chi_\text{mm}^{yy}$. Moreover, the relations that will be derived in what follows are also valid for isotropic structures since isotropy is a particular case of birefringence, which occurs when $\chi_\text{ee}^{xx}=\chi_\text{ee}^{yy}$ and $\chi_\text{mm}^{xx}=\chi_\text{mm}^{yy}$.

To obtain the dispersion relations of birefringent metasurfaces, we set all the susceptibility components in~\eqref{eq:MatAB} to zero except for the four terms mentioned above and solve the eigenvalue problem~\eqref{eq:Eig2}. This yields a set of four eigenvalues, four associated eigenvectors and their corresponding propagation constants, which are provided in Table~\ref{tab:bir}.
\begin{table}[h!]
\centering
\begin{tabular}{c|c|cN}
\hline
Eigenvalues,~$k_z$ & Eigenvectors,~$\ve{x}^\text{T}$ & Propagation constants,~$k_x$  &\\[15pt]
 \hline
$\frac{2j}{\chi_\text{mm}^{xx}}$ & \begin{tabular}[c]{@{}c@{}} $(-1,0,1,0)$\\ TE, symmetric, $\omega^-$ \end{tabular} & $\pm \frac{\sqrt{4+k^2\chi_\text{mm}^{2xx}}}{\chi_\text{mm}^{xx}}$ & \\[15pt]
 \hline
$\frac{2j}{\chi_\text{ee}^{xx}}$ & \begin{tabular}[c]{@{}c@{}} $(0,-1,0,1)$ \\ TM, symmetric, $\omega^-$ \end{tabular} & $\pm \frac{\sqrt{4+k^2\chi_\text{ee}^{2xx}}}{\chi_\text{ee}^{xx}}$ & \\[15pt]
 \hline
$-\frac{1}{2}jk^2\chi_\text{ee}^{yy}$ & \begin{tabular}[c]{@{}c@{}} $(1,0,1,0)$ \\ TE, asymmetric, $\omega^+$ \end{tabular} & $\pm\frac{1}{2}k\sqrt{4+k^2\chi_\text{ee}^{2yy}}$ & \\[15pt]
 \hline
$-\frac{1}{2}jk^2\chi_\text{mm}^{yy}$ & \begin{tabular}[c]{@{}c@{}} $(0,1,0,1)$ \\ TM, asymmetric, $\omega^+$  \end{tabular} & $\pm\frac{1}{2}k\sqrt{4+k^2\chi_\text{mm}^{2yy}}$ & \\[15pt]
\hline
\end{tabular}
  \caption{Dispersion relations of birefringent metasurfaces.}
  \label{tab:bir}
\end{table}

We see that a birefringent metasurface may support both TE and TM modes, which are decoupled from each other since this kind of structure does not induce rotation of polarization. Due to the symmetry of the environment, the absolute value of the amplitude of the surface waves on both sides of the metasurface is the same. However, the interaction of the top and bottom surface waves at the metasurface results in a symmetric and asymmetric mode splitting, where the symmetric mode corresponds to in-phase tangential electric fields and the asymmetric mode corresponds to out-of-phase tangential electric fields. For a given value of the propagation constant $k_x$, the symmetric mode is associated with a frequency $\omega^-$ lower than the frequency $\omega^+$ of the asymmetric mode.

In order to demonstrate the validity of the dispersion relations provided in Table~\ref{tab:bir}, we will use them to evaluate the dispersion of some relatively simple structures. Before discussing the case of a birefringent metasurface, we will first consider the case of thin isotropic slabs for which there exists exact dispersion relations directly derived using Maxwell equations and by applying the boundary conditions at both interfaces.

In the case of a slab surrounded by vacuum on both sides, the dispersion relations of the symmetric and asymmetric modes are respectively given by~\cite{raether1988surface}
\begin{subequations}\label{eq:slab2Modes}
\begin{align}
\omega^- : ~& \epsilon_{rs} k_{z} + k_{sz} \coth{\left(\frac{k_{sz}d}{2j}\right)}  = 0,\\
\omega^+ : ~& \epsilon_{rs} k_{z} + k_{sz} \tanh{\left(\frac{k_{sz}d}{2j}\right)}  = 0,
\end{align}
\end{subequations}
where $d$ is the thickness of the slab, $\epsilon_{rs}$ is its relative permittivity, and $k_{z}$ and $k_{sz}$ are the normal wavenumbers in vacuum and in the slab, respectively, The corresponding dispersion diagram of the slab can now be obtained by numerically solving relations~\eqref{eq:slab2Modes}. In what follows, we will compare the dispersion diagram found using relations~\eqref{eq:slab2Modes} to that found from the relations in Table~\ref{tab:bir}.

To compute the dispersion diagram using the GSTCs method, we first have to obtain the effective surface susceptibilities of the slab. This operation of homogenization may be easily achieved by relating the susceptibilities to the normal reflection and transmission coefficients, as explained in Appendix~\ref{app:Sparam}. In our case, the susceptibilities are directly obtained using~\eqref{eq:reducedSys} as
\begin{subequations}\label{eq:Xeexx}
\begin{align}
\chi_\text{ee}^{aa} &= \frac{2j}{k}\left(\frac{S_{21}^{aa} + S_{11}^{aa} - 1}{S_{21}^{aa} + S_{11}^{xx} + 1}\right), \\
\chi_\text{mm}^{bb} &= \frac{2j}{k}\left(\frac{S_{21}^{aa} - S_{11}^{aa} - 1}{S_{21}^{aa} - S_{11}^{aa} + 1}\right),
\end{align}
\end{subequations}
where $aa, bb = \{xx, yy\}$ and $S_{11}^{aa}$ and $S_{21}^{aa}$ are the slab reflection and transmission coefficients for $a$-polarized waves, respectively. Since the slab is isotropic, we have that $\chi_\text{ee}^{xx} = \chi_\text{ee}^{yy}$ and $\chi_\text{mm}^{xx} = \chi_\text{mm}^{yy}$, as said previously.

Note that even though the slab has a relative permeability $\mu_{rs} = 1$, it does not mean that its corresponding effective magnetic susceptibility is zero. Indeed, due to the nonzero thickness of the slab, there is a variation of the electromagnetic field distribution within it. Within the zero-thickness model of the GSTCs, this variation of the field distribution inside the slab can only be modeled by a combination of nonzero electric and magnetic susceptibilities.

\begin{figure}[h!]
\centering
\subfloat[]{\label{fig:SlabDisp1}
\includegraphics[width=0.9\columnwidth]{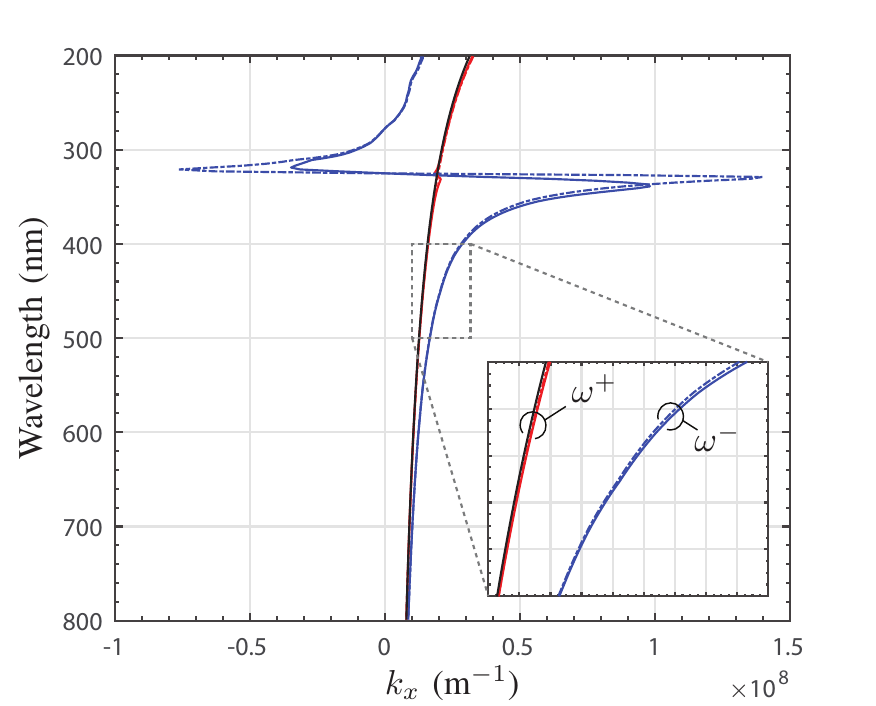}
}\\
\subfloat[]{\label{fig:SlabDisp2}
\includegraphics[width=0.9\columnwidth]{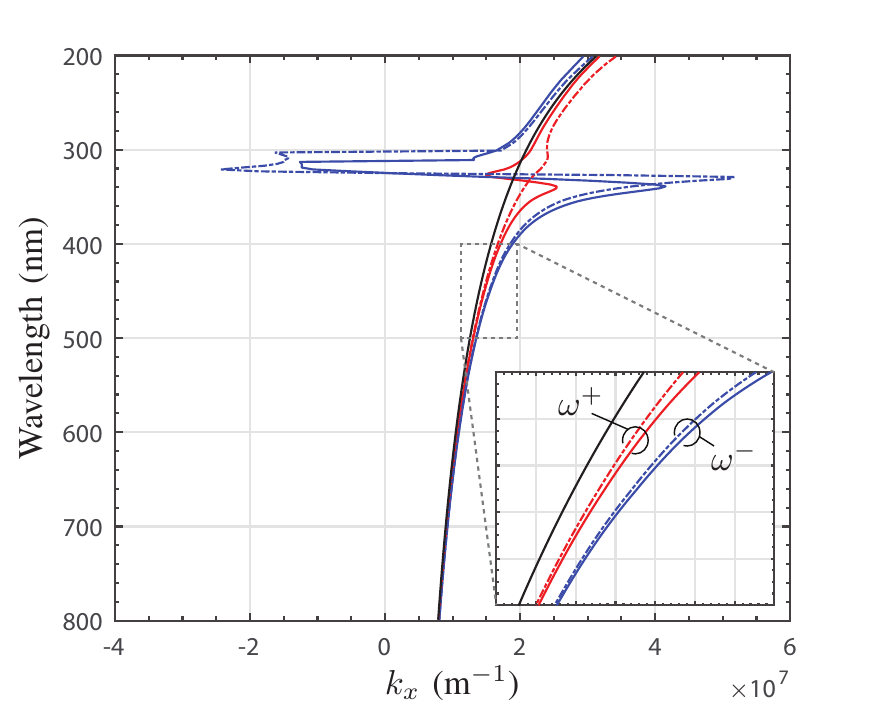}
}
\caption{Dispersion diagrams for silver slabs with thicknesses (a)~$d=20$~nm and (b)~$d=60$~nm. The black curve is the light line. The dashed curves correspond to the GSTCs method, while the solid curves correspond to the exact solutions.}
\label{fig:SilverSlabDisp}
\end{figure}

\begin{figure}[h!]
\centering
\subfloat[]{\label{fig:GoldSlabDisp1}
\includegraphics[width=0.9\columnwidth]{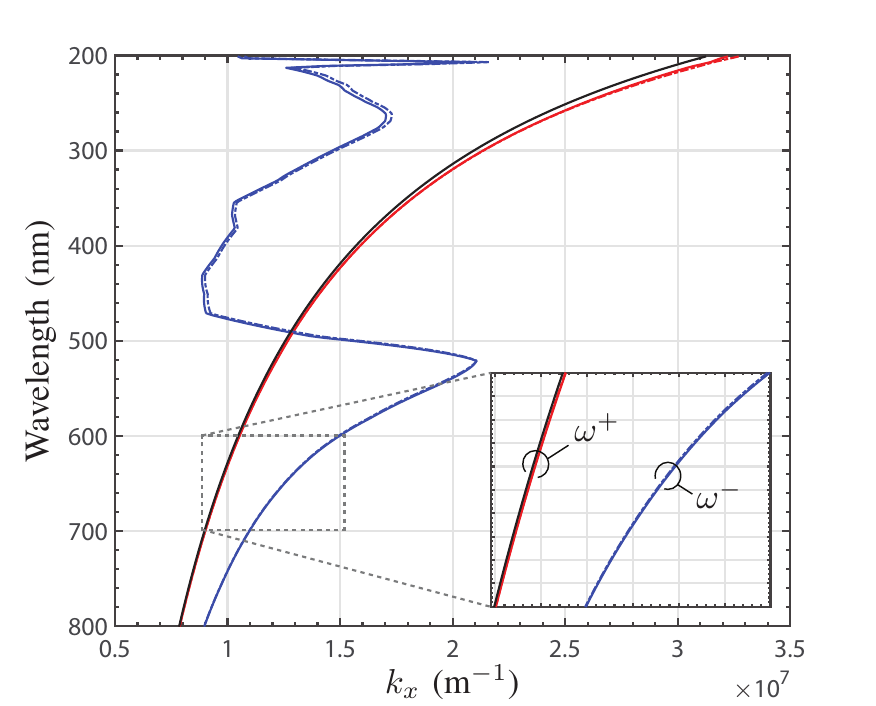}
}\\
\subfloat[]{\label{fig:GoldSlabDisp2}
\includegraphics[width=0.9\columnwidth]{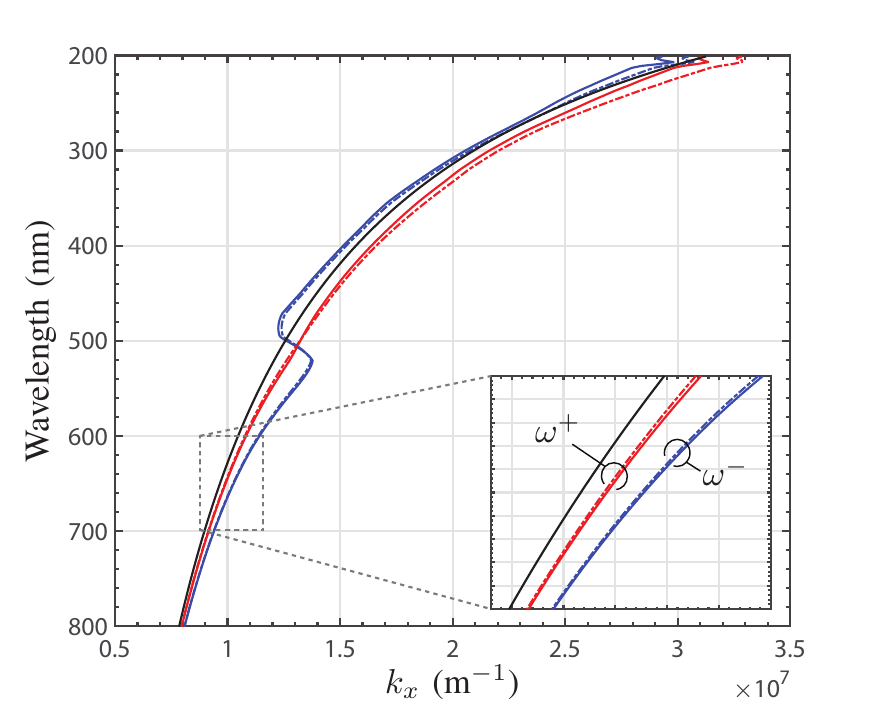}
}
\caption{Dispersion diagrams for gold slabs with thicknesses (a)~$d=20$~nm and (b)~$d=60$~nm. The black curve is the light line. The dashed curves correspond to the GSTCs method, while the solid curves correspond to the exact solutions.}
\label{fig:GoldSlabDisp}
\end{figure}

The reflection and transmission coefficients of the slab are easily found analytically. Then, substituting relations~\eqref{eq:Xeexx} in the dispersion relations in Table~\ref{tab:bir}, we compute the dispersion diagrams of two different metallic slabs. The resulting dispersion diagrams are plotted in Fig.~\ref{fig:SilverSlabDisp}, for a slab made out of silver, and in Fig.~\ref{fig:GoldSlabDisp}, for a slab made out of gold. For these two slabs, we consider a thickness of $d=20$~nm and $d=60$~nm, and the material parameters are found from~\cite{JohnsonChristy}. In the two figures, the dashed curves correspond to the GSTCs method, while the solid curves correspond to the exact solutions.

Overall, there is a very good agreement between the two methods at the exception of the region near the plasmon resonance where we notice some discrepancies. This may be explained by the fact that, near resonance, the fields are more confined within the slabs thus increasing the role played by the normal susceptibility components that we have initially decided to ignore in~\eqref{eq:GSTCs}. This suggests that it may be possible to achieve an even better result by taking into account the presence of normal susceptibility components, but the developments required to consider these susceptibilities are beyond the scope of this work.

\begin{figure}[h!]
\centering
\subfloat[]{\label{fig:ComsolIso1}
\includegraphics[width=0.48\columnwidth]{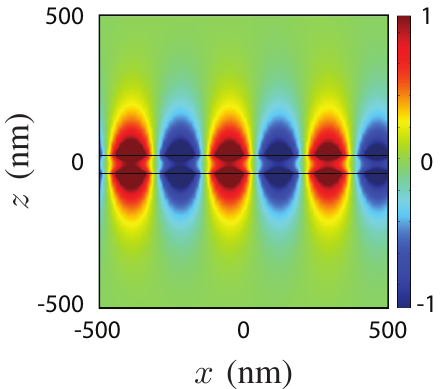}
}
\subfloat[]{\label{fig:ComsolIso2}
\includegraphics[width=0.48\columnwidth]{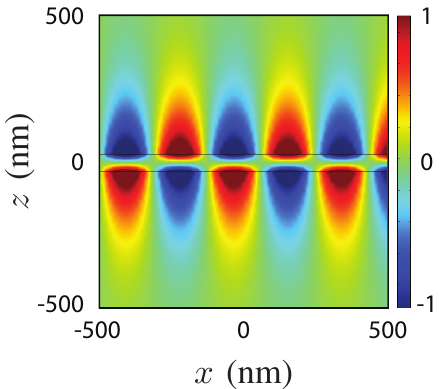}
}
\caption{Simulated real part of $E_x$ for surface waves, with (a)~symmetric and (b) asymmetric field distributions, propagating on a 60-nm thick silver slab at $\lambda = 400$~nm.}
\label{fig:ComsolIso}
\end{figure}

We note that the two metallic slabs support the propagation of both symmetric (blue curves) and asymmetric (red curves) TM modes. An illustration of these two modes is presented in Fig.~\ref{fig:ComsolIso} for comparison. As a general observation, the thinner the slab, the more decoupled are the two modes, with the asymmetric mode converging more and more towards the light line for very thin slabs. In contrast, when the thickness of the slabs increases, the two modes converges and finally completely overlap for very thick slabs.\\

In terms of synthesis, the expressions provided in Table~\ref{tab:bir} may be directly used to find the susceptibilities required to achieve the propagation of a desired surface-wave mode. Let us for instance consider the synthesis of a metasurface supporting a TM surface-wave mode with a specified propagation constant $k_{x,\text{spec}}>k$ and associated normal wavenumber $k_{z,\text{spec}}^2 = k^2 - k_{x,\text{spec}}^2$. Using the eigenvalues provided in the table, we simple have that $\chi_\text{ee}^{xx} = 2j/k_{z,\text{spec}}$, for a symmetric mode, and  $\chi_\text{mm}^{yy} = 2jk_{z,\text{spec}}/k^2$, for an asymmetric mode, respectively. Implementing either of these two susceptibilities will allow one to achieve TM surface-wave propagation of either a symmetric or an asymmetric mode.

At this point, one may ask the question: what happens if a metasurface is implemented so as to simultaneously have that $\chi_\text{ee}^{xx} = 2j/k_{z,\text{spec}}$ and $\chi_\text{mm}^{yy} = 2jk_{z,\text{spec}}/k^2$ for the same value of $k$? In that case, the metasurface is able to simultaneously support the propagation of both a symmetric and an asymmetric TM surface-wave modes at the same frequency. Therefore, by constructive or destructive superposition of these two modes, the surface wave will only exist on the bottom or on the top of the metasurface, respectively, which corresponds to the results already discussed in~\cite{Achouri2016b}. Note that this peculiar effect is purely based on the interference between the symmetric and asymmetric modes. Consequently, if they are not exactly in-phase or out-of-phase, then the surface wave will partially exist on both sides of the metasurface. In the next section, we will see a method which allows one to completely forbid the existence of a surface wave on one side of the metasurface; a phenomenon that is not based on the superposition of two surface waves like the one presented here.\\

\begin{figure}[ht]
\begin{center}
\includegraphics[width=0.8\columnwidth]{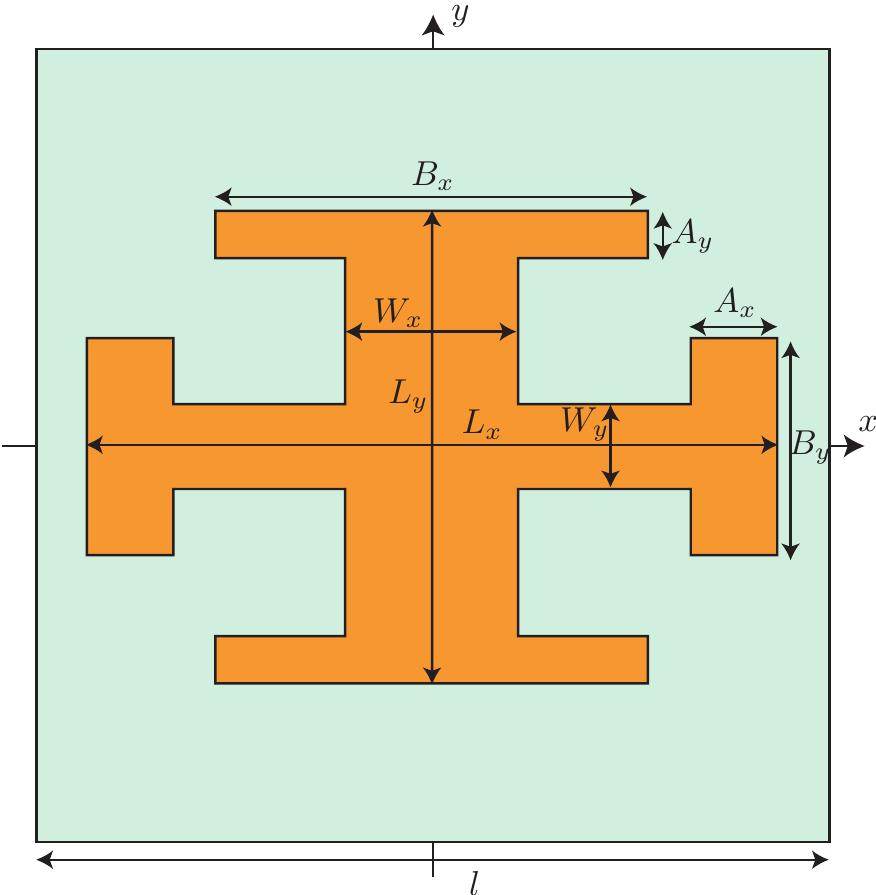}
\caption{Representations of a generic Jerusalem cross with the dimensions that can be modified. The metal is copper and the dielectric slab is a Rogers RO3003 substrate with $\epsilon_r=3$ and $\tan \delta = 0.001$.}
\label{Fig:metallicUnitCell}
\end{center}
\end{figure}

Now that the validity of the GSTCs dispersion model has been established, we will use it to compute the dispersion diagram of a birefringent metasurface structure. We have selected a unit cell which was previously used to realize a microwave surface-wave guiding metasurface~\cite{Achouri2016b}. The unit cell structure is made of three metallic layers separated by two dielectric spacers. Each of the metallic layer takes the shape of a Jerusalem cross as shown in Fig.~\ref{Fig:metallicUnitCell}. The structure is designed for a central frequency of $f=10$~GHz. The lateral dimension of the unit cell is $l=6$~mm and its thickness is $d=3.04$~mm.
This structure was numerically optimized, using an eigenmode solver, so as to support the propagation of surface waves~\cite{Achouri2016b}. The resulting optimal dimensions of the Jerusalem crosses are reported in Table~\ref{TableGuidSW}.
\begin{table}[h!]
\centering
\begin{tabular}{cccccccccc}  \toprule
 Layer  & Lx   & Ly & Wx  & Wy   & Ax   & Ay   & Bx   & By   \\\midrule
OL      & 5.5    & 4  & 0.5 & 1 & 0.5  & 0.5  & 2.25    & 2.25 \\
ML      & 5 & 5  & 0.25 & 1 & 0.5 & 0.5  & 3.75 & - \\\bottomrule
\end{tabular}
\caption{Geometrical dimensions (in mm) of a Jerusalem cross. OL denotes the outer layers and ML the middle layer.}\label{TableGuidSW}
\end{table}

\begin{figure}[h!]
\centering
\includegraphics[width=0.9\columnwidth]{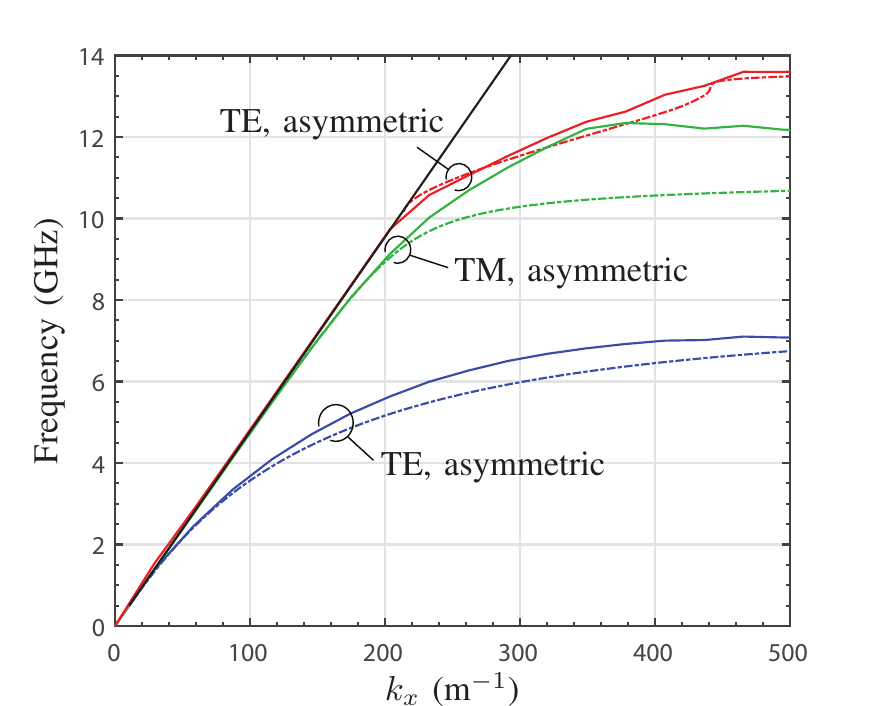}
\caption{Dispersion diagram of a symmetric metasurface unit cell. The black curve is the light line. The dashed curves correspond to the GSTCs method, while the solid curves correspond to the full-wave numerical solutions.}
\label{fig:UnitCellDisp}
\end{figure}

The dispersion diagram computed using the eigenmode solver is now compared to the approximate dispersion diagram found using the GSTCs method, where the scattering parameters of the unit cell were obtained by full-wave simulations. The resulting dispersion diagrams are plotted for comparison in Fig.~\ref{fig:UnitCellDisp}.
Due to the complex geometry of the unit cell, this metasurface supports several modes among which only three are plotted in the figure. Interestingly, we note that this kind of structure supports both TE and TM asymmetric modes.

The comparison between the dispersion curves obtained with the eigenmode solver and the GSTCs method reveals an overall good agreement, at least for the TE modes. Below 9~GHz, the two methods yields almost identical results for the TM mode, however a non-negligible discrepancy is present near the resonance. As explained previously, this discrepancy is likely to stem from the presence of normal susceptibility components that are ignored in the GSTCs method. The reason that the discrepancy is more important for the TM mode than the TE modes is explained by the fact that this type of unit cell structure exhibits a normal electric susceptibility, $\chi_\text{ee}^{zz}$, that is much more important than its magnetic counterpart, $\chi_\text{mm}^{zz}$, due to the tendency of the field component $E_z$ to be strongly confined between the metallic layers near the resonance. Moreover, this effect is maximized in the case of a TM asymmetric mode since, for this mode, the $E_z$ components of the top and bottom surface waves are in-phase within the structure, which magnifies the effect of $\chi_\text{ee}^{zz}$ and thus increases the discrepancy.

\subsection{Dispersion of Omega-Type Metasurfaces}

Omega-type metasurfaces are a particular class of structures that induce no rotation of polarization and that are structurally asymmetric in their longitudinal (here $z$) direction~\cite{Tretyakov1998}. Due to their asymmetry, they exhibit reflection coefficients that are different for forward and backward illuminations, which has been recently leveraged to realize refractive metasurfaces without spurious diffraction~\cite{Asadchy2016,7506314,8259235}. Their asymmetry may thus be seen as a source of additional degrees of freedom to control the propagation of surface waves. Specifically, one may ask the question: can this asymmetry be leveraged to arbitrarily control the amplitude of the TE or TM surface-wave modes on both sides of the metasurface? To answer this question, we will now solve the eigenvalue problem~\eqref{eq:Eig2} for the case of omega-type metasurfaces.

An omega-type metasurface is a bianisotropic structure, which possesses the following nonzero susceptibility components: $\chi_\text{ee}^{xx},\chi_\text{ee}^{yy},\chi_\text{mm}^{xx},\chi_\text{mm}^{yy},\chi_\text{em}^{xy},\chi_\text{em}^{yx},\chi_\text{me}^{xy}$ and $\chi_\text{me}^{yx}$. For simplicity, we restrict our attention to the case of reciprocal metasurfaces, which relates some of the susceptibility components to each other through~\eqref{eq:reciprocity}.
We may now solve the eigenvalue problems~\eqref{eq:Eig2} for TE and TM surface-wave propagation. The resulting eigenvalues and eigenvectors are given in Table~\ref{tab:omega}, while the corresponding propagation constants are not provided for the sake of briefness but may be straightforwardly found by solving $k^2 = k_x^2+k_z^2$ for each eigenvalue.
\begin{table}[h!]
\centering
\begin{tabular}{c|cN}
\hline
Eigenvalues,~$k_z$ & Eigenvectors,~$\ve{x}^\text{T}$  &\\[15pt]
 \hline
$\frac{k^2\left(\chi_\text{em}^{2yx}+\chi_\text{ee}^{yy}\chi_\text{mm}^{xx}\right)-4+C_\text{TE}}{4j\chi_\text{mm}^{xx}}$
 & $\left(\frac{4jk\chi_\text{em}^{yx} + C_\text{TE}}{k^2\left(\chi_\text{em}^{2yx}+\chi_\text{ee}^{yy}\chi_\text{mm}^{xx}\right)+4},0,1,0\right)$ &  \\[20pt]
 \hline
$\frac{k^2\left(\chi_\text{em}^{2yx}+\chi_\text{ee}^{yy}\chi_\text{mm}^{xx}\right)-4-C_\text{TE}}{4j\chi_\text{mm}^{xx}}$
 &  $\left(\frac{4jk\chi_\text{em}^{yx} - C_\text{TE}}{k^2\left(\chi_\text{em}^{2yx}+\chi_\text{ee}^{yy}\chi_\text{mm}^{xx}\right)+4},0,1,0\right)$  & \\[20pt]
 \hline
$\frac{k^2\left(\chi_\text{em}^{2xy}+\chi_\text{ee}^{xx}\chi_\text{mm}^{yy}\right)-4+C_\text{TM}}{4j\chi_\text{ee}^{xx}}$
 &  $\left(0,\frac{4jk\chi_\text{em}^{xy} + C_\text{TM}}{k^2\left(\chi_\text{em}^{2xy}+\chi_\text{ee}^{xx}\chi_\text{mm}^{yy}\right)+4},0,1\right)$  & \\[20pt]
 \hline
$\frac{k^2\left(\chi_\text{em}^{2xy}+\chi_\text{ee}^{xx}\chi_\text{mm}^{yy}\right)-4-C_\text{TM}}{4j\chi_\text{ee}^{xx}}$
 &  $\left(0,\frac{4jk\chi_\text{em}^{xy} - C_\text{TM}}{k^2\left(\chi_\text{em}^{2xy}+\chi_\text{ee}^{xx}\chi_\text{mm}^{yy}\right)+4},0,1\right)$  &  \\[20pt]
\hline
\multicolumn{2}{c}{\begin{tabular}[c]{@{}c@{}}$C_\text{TE} = \sqrt{k^4\chi_\text{em}^{4yx}+2k^2\chi_\text{em}^{2yx}(k^2\chi_\text{ee}^{yy}\chi_\text{mm}^{xx}-4)+(k^2\chi_\text{ee}^{yy}\chi_\text{mm}^{xx}+4)^2}$\\ $C_\text{TM} = \sqrt{k^4\chi_\text{em}^{4xy}+2k^2\chi_\text{em}^{2xy}(k^2\chi_\text{ee}^{xx}\chi_\text{mm}^{yy}-4)+(k^2\chi_\text{ee}^{xx}\chi_\text{mm}^{yy}+4)^2}$ \end{tabular}}  & \\[25pt]
\hline
\end{tabular}
  \caption{Dispersion relations for omega-type metasurfaces. The propagation constants, $k_x$, are easily calculated by solving $k^2= k_z^2+k_x^2$.}
  \label{tab:omega}
\end{table}

As for the case of birefringent metasurfaces discussed in the previous section, the expressions provided in this table may be used to perform a metasurface synthesis. Specifically, we see from the expressions of the eigenvectors, that it is possible to control the amplitude of the TE/TM modes that propagates on both sides of the metasurface. These expressions notably suggest that, for a specific TE or TM mode, it should be possible to achieve surface-wave propagation only on one side of the metasurface, while completely forbidding the existence of surface waves of the same mode on the other side.

In order to demonstrate that this is indeed possible, we now derive the susceptibilities that allow the propagation of a TM surface wave only on the bottom side of an omega-type metasurface. This specification implies the following two conditions: 1)~the vector $\ve{x}^\text{T} = (0,1,0,0)$, corresponding to TM surface-wave propagation only on the bottom side of the metasurface, is an eigenvector of~\eqref{eq:Eig2}, and 2)~the vector $\ve{x}^\text{T} = (0,0,0,1)$, corresponding to TM surface-wave propagation of the top side of the metasurface, is not an eigenvector of that system. These two conditions lead to a system of equations that may be easily solved and yields the following susceptibilities and propagation constant:
\begin{equation}\label{eq:BottomTMSW}
\chi_\text{mm}^{yy} = 0,\quad \chi_\text{em}^{xy} = \frac{2j}{k},\quad k_x = \pm \frac{\sqrt{16+k^2\chi_\text{ee}^{2xx}}}{\chi_\text{ee}^{xx}},
\end{equation}
where $\chi_\text{ee}^{xx}$ is left as a free parameter that may be used to achieve the desired propagation constant. Note that all the other susceptibilities not defined in~\eqref{eq:BottomTMSW} are also left as free parameters since the TM wave does not interact with them.

\begin{figure}[h!]
\centering
\includegraphics[width=0.5\columnwidth]{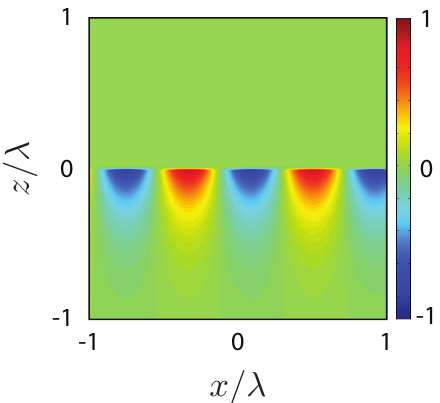}
\caption{FDFD simulated real part of $E_x$ of a bianisotropic metasurface allowing the propagation of a TM surface only on its bottom side.}
\label{fig:BianiSW}
\end{figure}

We have performed a numerical simulation to verify that the susceptibilities in~\eqref{eq:BottomTMSW} allow the propagation of a TM surface only on the bottom of the metasurface and prevents the propagation of TM surface waves on its top. We have arbitrarily specified that $k_x = 1.2 k$, which allows us to define the value of $\chi_\text{ee}^{xx}$. The simulation is performed using an homemade finite-difference frequency-domain (FDFD) scheme~\cite{Vahabzadeh2016,vahabzadeh2017computational}. The resulting real part of $E_x$ is plotted in Fig.~\ref{fig:BianiSW}.
As can be seen, the metasurface, which is implemented as a zero-thickness sheet at $z=0$, only supports a surface wave on its bottom side.

Following the same procedure, we provide in Table~\ref{tab:SingleSW} the conditions for unilateral surface-wave propagation for the four different combinations of TE/TM polarizations and top/bottom propagations.
\begin{table}[h!]
\centering
\begin{tabular}{c|c|cN}
\hline
Description & Susceptibilities & Propagation constans, $k_x$ &\\[15pt]
 \hline
TM wave at $z=0^-$ & \begin{tabular}[c]{@{}c@{}} $\chi_\text{mm}^{yy} = 0$ \\ $\chi_\text{em}^{xy} = \frac{2j}{k}$ \end{tabular} & $k_x = \pm \frac{\sqrt{16+k^2\chi_\text{ee}^{2xx}}}{\chi_\text{ee}^{xx}}$ &\\[20pt]
 \hline
TM wave at $z=0^+$ & \begin{tabular}[c]{@{}c@{}} $\chi_\text{mm}^{yy} = 0$ \\ $\chi_\text{em}^{xy} = -\frac{2j}{k}$ \end{tabular} & $k_x = \pm \frac{\sqrt{16+k^2\chi_\text{ee}^{2xx}}}{\chi_\text{ee}^{xx}}$ &\\[20pt]
 \hline
TE wave at $z=0^-$ & \begin{tabular}[c]{@{}c@{}} $\chi_\text{ee}^{yy} = 0$ \\ $\chi_\text{em}^{yx} = \frac{2j}{k}$ \end{tabular} & $k_x = \pm \frac{\sqrt{16+k^2\chi_\text{mm}^{2xx}}}{\chi_\text{mm}^{xx}}$ &\\[20pt]
 \hline
TE wave at $z=0^+$ & \begin{tabular}[c]{@{}c@{}} $\chi_\text{ee}^{yy} = 0$ \\ $\chi_\text{em}^{yx} = -\frac{2j}{k}$ \end{tabular} & $k_x = \pm \frac{\sqrt{16+k^2\chi_\text{mm}^{2xx}}}{\chi_\text{mm}^{xx}}$ &\\[20pt]
\hline
\end{tabular}
  \caption{Synthesis conditions for unilateral surface-wave propagation on omega-type metasurfaces.}
  \label{tab:SingleSW}
\end{table}

\section{Dispersion in an Asymmetric Environment}
\label{sec:asym}

We are now interested in the case where the media on both sides of the metasurface are different from each other, which implies that $\eta_1 \neq \eta_2$, $k_1 \neq k_2$ and $k_{1z} \neq k_{2z}$. As a consequence, the latter inequality prevents the reduction of the GSTCs model to the convenient eigenvalue problem of Eq.~\eqref{eq:Eig2} since this equation is only valid when $k_{1z} = k_{2z}$. Nevertheless, it is still possible to obtain the dispersion relations in this asymmetric environment by directly solving Eqs.~\eqref{eq:GeneralEq} and by noting that, even if the normal wavenumbers are different from each other, the propagation constant, $k_x$, must be the same on both sides of the metasurface by phase matching. Therefore, one may obtain the dispersion relations by respectively substituting $k_{1z}$ and $k_{2z}$ by $k_{1z} = \pm\sqrt{k_1^2-k_x^2}$ and $k_{2z} = \pm\sqrt{k_2^2-k_x^2}$ and solving Eqs.~\eqref{eq:GeneralEq} for $k_x$. The general solution to this system of equations is particularly cumbersome and is thus not presented here. In what follows, we will rather consider a simplified but still experimentally very relevant situation, which is that of an isotropic thin structure surround by two different media.

\begin{figure}[h!]
\centering
\includegraphics[width=0.9\columnwidth]{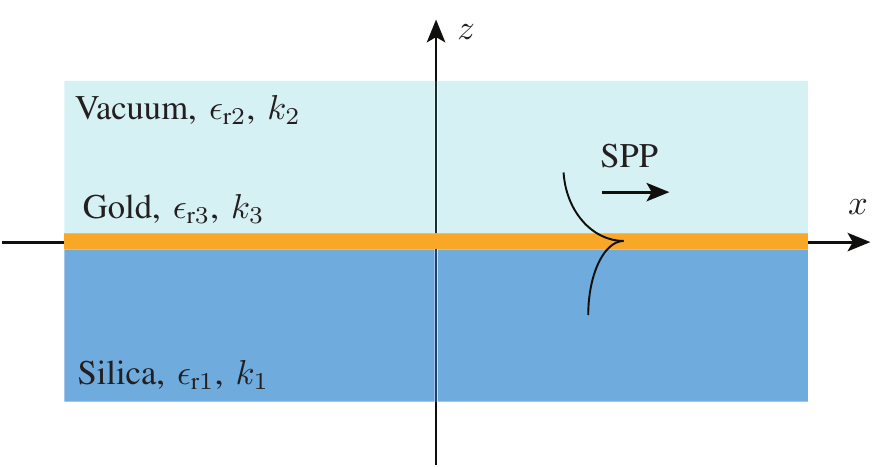}
\caption{Asymmetric environment consisting of a 20-nm gold layer surrounded by vacuum and silica, which supports the propagation of a surface plasmon.}
\label{fig:asym}
\end{figure}

Let us consider the two-dimensional asymmetric environment depicted in Fig.~\ref{fig:asym}.
In order to find the dispersion relations of this thin metallic slab, we first have to obtain its effective susceptibilities. As before, this may be achieved by using the relationship between the effective susceptibilities and the scattering parameters discussed in Appendix~\ref{app:Sparam}. However, due to the asymmetry of this environment, it is not possible to use the simple relations~\eqref{eq:Xeexx}, as done in the case of the symmetric environment. Indeed, these relations can only be used if the slab can be modeled by an (an)isotropic zero-thickness sheet, which is not the case of the slab in Fig.~\ref{fig:asym}. This is because an (an)isotropic sheet exhibits the \emph{same} scattering parameters irrespectively of the direction of the illumination, while the structure in Fig.~\ref{fig:asym} exhibits \emph{different} reflection coefficients for forward and backward illumination. Therefore, the correct way to model this slab is to replace it by a zero-thickness bianisotropic sheet, whose effective susceptibilities are found using~\eqref{eq:reducedSys} along with~\eqref{eq:DelAv} in Appendix~\ref{app:Sparam}. Note that we have here specifically chosen for simplicity a structure that is isotropic and which thus does not induce rotation of polarization and/or coupling between TE and TM modes. Consequently, the system~\eqref{eq:reducedSys} may be significantly reduced and the susceptibilities may be found from the scattering parameters directly using~\eqref{eq:BianiStoX} in Appendix~\ref{app:Sparam}.

Since we may treat TE and TM modes independently from each other, the system~\eqref{eq:GeneralEq} splits into two separates systems, one for each mode. We thus have
\begin{subequations}\label{eq:BianiSys}
\begin{equation}\label{eq:BianiSysTM}
\te{\chi}_\text{TM}\cdot
\begin{pmatrix}
  A_\text{TM}^{0^-} \\
  A_\text{TM}^{0^+}
\end{pmatrix}=0,
\end{equation}
\begin{equation}\label{eq:BianiSysTE}
\te{\chi}_\text{TE}\cdot
\begin{pmatrix}
  A_\text{TE}^{0^-} \\
  A_\text{TE}^{0^+}
\end{pmatrix}=0,
\end{equation}
\end{subequations}
where the matrices $\te{\chi}_\text{TM}$ and $\te{\chi}_\text{TE}$ are given below in Eqs.~\eqref{eq:ChiMatBiani}. These two sets of equations cannot be cast into the convenient eigenvalue formulation that was used in~\eqref{eq:Eig2}. However, they may still be solved to extract the propagation constant, $k_x$. To do so, we first substitute the $k_z$ wavenumbers by $k_{1z} = \pm\sqrt{k_1^2-k_x^2}$ and $k_{2z} = \pm\sqrt{k_2^2-k_x^2}$, and use the fact that we are not interested in the exact values of the TE and TM modes amplitude but rather on the ratios between the top and bottom waves, i.e. $A_\text{TM}^{0^+}/A_\text{TM}^{0^-}$ and $A_\text{TE}^{0^+}/A_\text{TE}^{0^-}$. Accordingly, we arbitrarily chose that $A_\text{TM}^{0^+} = A_\text{TE}^{0^+} = 1$, which reduces~\eqref{eq:BianiSys} to two systems of two equations in two unknowns.
\begin{figure}[h!]
\centering
\includegraphics[width=1\columnwidth]{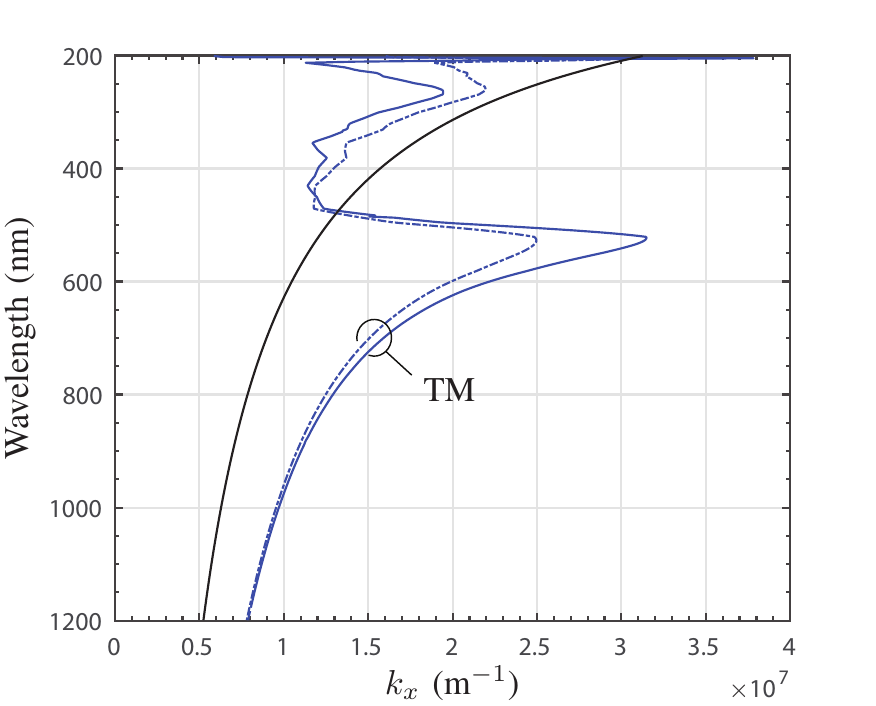}
\caption{Dispersion diagram of a 20-nm gold layer surrounded by vacuum and silica. The black curve is the light line. The dashed curve corresponds to the GSTCs method, while the solid curve corresponds to the exact solution.}
\label{fig:AsymSlab}
\end{figure}

For the problem of Fig.~\ref{fig:asym}, the metal layer only supports TM modes and we thus only have to solve~\eqref{eq:BianiSysTM} for the surface wave propagation constants. For such a simple structure, there exists an exact solution for the dispersion relation, which we will use to compare our model to. This exact dispersion relation reads~\cite{maier2007plasmonics}
\begin{equation}\label{eq:ExactAsym}
e^{-2k_1d} = \left( \frac{k_3/\epsilon_{\text{r}3}+k_1/\epsilon_{\text{r}1}}{k_3/\epsilon_{\text{r}3}-k_1/\epsilon_{\text{r}1}}\right)\cdot\left(\frac{k_3/\epsilon_{\text{r}3}+k_2/\epsilon_{\text{r}2}}{k_3/\epsilon_{3\text{r}}-k_2/\epsilon_{\text{r}2}}\right),
\end{equation}
where $d$ is the thickness of the metal layer and $\epsilon_{\text{r}i}$ and $k_i^2 = k_{ix}^2 + k_{iz}^2$ respectively refer to the relative permittivity and the wavenumber of each medium in Fig.~\ref{fig:asym}.

The resulting exact and approximate dispersion curves of the metal layer of Fig.~\ref{fig:asym} are plotted in Fig.~\ref{fig:AsymSlab} in solid and dashed line, respectively. Note that we have assumed that the permittivity of silica is constant over the specified bandwidth and equal to $\epsilon_{\text{r}1}= 2.09$, while the dispersion of gold is found from~\cite{JohnsonChristy}.

As for the other cases previously discussed, we see a good agreement between the exact TM mode of the structure and the TM mode predicted by the GSTCs method. This example thus demonstrates the capabilities of the GSTCs method to approximate the dispersion of bianisotropic metasurfaces in both symmetric and asymmetric environments.

\begin{subequations}\label{eq:ChiMatBiani}
\begin{floatEq}
\begin{equation}\label{eq:ChiMatTM}
\te{\chi}_\text{TM}=
\begin{pmatrix}
   -2+jk\chi_\text{em}^{xy}-\frac{jk_{1z}\eta_1\omega\epsilon\chi_\text{ee}^{xx}}{k_1} & 2+jk\chi_\text{em}^{xy} +\frac{jk_{2z}\eta_2\omega\epsilon\chi_\text{ee}^{xx}}{k_2} \\
    -j\omega\mu\chi_\text{mm}^{yy}-\frac{k_{1z}\eta_1(2-jk\chi_\text{me}^{yx})}{k_1}  & -j\omega\mu\chi_\text{mm}^{yy} -\frac{k_{2z}\eta_2(2+jk\chi_\text{me}^{yx})}{k_2}
\end{pmatrix}.
\end{equation}
\begin{equation}\label{eq:ChiMatTE}
\te{\chi}_\text{TE}=
\begin{pmatrix}
   j\omega\epsilon\chi_\text{ee}^{yy}+\frac{k_{1z}(2+jk\chi_\text{em}^{yx})}{k_1\eta_1} & j\omega\epsilon\chi_\text{ee}^{yy}+\frac{k_{2z}(2-jk\chi_\text{em}^{yx})}{k_2\eta_2} \\
  -2-jk\chi_\text{me}^{xy}-\frac{jk_{1z}\omega\mu\chi_\text{mm}^{xx}}{k_1 \eta_1}  & 2-jk\chi_\text{me}^{xy}+\frac{jk_{2z}\omega\mu\chi_\text{mm}^{xx}}{k_2 \eta_2}
\end{pmatrix}.
\end{equation}
\end{floatEq}
\end{subequations}

\section{Conclusion}

In this work, we have established a connection between the effective susceptibilities of bianisotropic metasurfaces and their capability of supporting the propagation of surface waves. Based on several examples, we have demonstrated the effectiveness of the GSTCs method to retrieve and approximate the dispersion of such structures. This work has also revealed that ignoring the presence of normal polarizations leads to a simplified system of equations, which yields dispersion curves that are in very good agreement with the expected response of these structures, at least away from the plasmon resonance. Near the resonance, the strong field confinement within the structures triggers the normal polarizations, which leads to an inaccuracy of the model in this particular frequency region.

In addition of providing a framework for computing the dispersion of bianisotropic metasurfaces, this work also provides the tools required to theoretically synthesize surface-wave guiding metasurfaces. Indeed, the analytical expressions relating the propagation constants of the surface-wave modes to the susceptibilities may be used to obtain the susceptibilities required to achieve a specified surface-wave propagation. In particular, we have shown that an omega-type bianisotropic metasurface can be synthesized to completely forbid the existence of a surface wave on one of its sides.

\section*{Acknowledgements}

We gratefully acknowledge funding from the European Research Council (ERC-2015-AdG-695206 Nanofactory).

\appendices

\section{Susceptibility Retrieval Method}
\label{app:Sparam}

In this appendix, we briefly summarize the main steps required to retrieve the susceptibilities of bianisotropic metasurfaces~\cite{achouri2017design}. The first step consists in casting equations~\eqref{eq:InvProb} into the following matrix system:
\begin{equation}
\label{eq:InvProbMatrix}
\begin{pmatrix}
\Delta H_y\\
\Delta H_x\\
\Delta E_y\\
\Delta E_x
\end{pmatrix}=
\begin{pmatrix}
\widetilde{\chi}_\text{ee}^{xx} & \widetilde{\chi}_\text{ee}^{xy} & \widetilde{\chi}_\text{em}^{xx} & \widetilde{\chi}_\text{em}^{xy}\\
\widetilde{\chi}_\text{ee}^{yx} & \widetilde{\chi}_\text{ee}^{yy} & \widetilde{\chi}_\text{em}^{yx} & \widetilde{\chi}_\text{em}^{yy}\\
\widetilde{\chi}_\text{me}^{xx} & \widetilde{\chi}_\text{me}^{xy} & \widetilde{\chi}_\text{mm}^{xx} & \widetilde{\chi}_\text{mm}^{xy}\\
\widetilde{\chi}_\text{me}^{yx} & \widetilde{\chi}_\text{me}^{yy} & \widetilde{\chi}_\text{mm}^{yx} & \widetilde{\chi}_\text{mm}^{yy}
\end{pmatrix}
\cdot
\begin{pmatrix}
E_{x,\text{av}}\\
E_{y,\text{av}}\\
H_{x,\text{av}}\\
H_{y,\text{av}}
\end{pmatrix},
\end{equation}
where the susceptibilities have been normalized according to the following convention:
\begin{equation}
\label{eq:conv}
\begin{split}
&
\begin{pmatrix}
\chi_\text{ee}^{xx} & \chi_\text{ee}^{xy} & \chi_\text{em}^{xx} & \chi_\text{em}^{xy}\\
\chi_\text{ee}^{yx} & \chi_\text{ee}^{yy} & \chi_\text{em}^{yx} & \chi_\text{em}^{yy}\\
\chi_\text{me}^{xx} & \chi_\text{me}^{xy} & \chi_\text{mm}^{xx} & \chi_\text{mm}^{xy}\\
\chi_\text{me}^{yx} & \chi_\text{me}^{yy} & \chi_\text{mm}^{yx} & \chi_\text{mm}^{yy}
\end{pmatrix}=\\
&\quad
=\begin{pmatrix}
\frac{j}{\omega\epsilon_0}\widetilde{\chi}_\text{ee}^{xx} & \frac{j}{\omega\epsilon_0}\widetilde{\chi}_\text{ee}^{xy} & \frac{j}{k_0}\widetilde{\chi}_\text{em}^{xx} & \frac{j}{k_0}\widetilde{\chi}_\text{em}^{xy}\\
-\frac{j}{\omega\epsilon_0}\widetilde{\chi}_\text{ee}^{yx} & -\frac{j}{\omega\epsilon_0}\widetilde{\chi}_\text{ee}^{yy} & -\frac{j}{k_0}\widetilde{\chi}_\text{em}^{yx} & -\frac{j}{k_0}\widetilde{\chi}_\text{em}^{yy}\\
-\frac{j}{k_0}\widetilde{\chi}_\text{me}^{xx} & -\frac{j}{k_0}\widetilde{\chi}_\text{me}^{xy} & -\frac{j}{\omega\mu_0}\widetilde{\chi}_\text{mm}^{xx} & -\frac{j}{\omega\mu_0}\widetilde{\chi}_\text{mm}^{xy}\\
\frac{j}{k_0}\widetilde{\chi}_\text{me}^{yx} & \frac{j}{k_0}\widetilde{\chi}_\text{me}^{yy} & \frac{j}{\omega\mu_0}\widetilde{\chi}_\text{mm}^{yx} & \frac{j}{\omega\mu_0}\widetilde{\chi}_\text{mm}^{yy}
\end{pmatrix}.
\end{split}
\end{equation}
Next, the matrix system~\eqref{eq:InvProbMatrix} is compactly rewritten as
\begin{equation}
\label{eq:reducedSys}
 \te{\Delta} = \widetilde{\te{\chi}}\cdot \te{A}_v,
\end{equation}
where $\te{\Delta}$, $\widetilde{\te{\chi}}$ and $\te{A}_v$ refer to the field differences, the normalized susceptibilities and the field averages, respectively. This system can easily be solved, by matrix inversion, to compute the susceptibilities in terms of specified fields. Conventionally, these fields are specified to be normally propagating incident, reflected and transmitted waves. In a very general situation, a normally incident $x$-polarized wave propagating in the positive $z$-direction may excite the metasurface such that it reflects and transmits $x$- and $y$-polarized waves. Accordingly, the corresponding electric fields of such an operation are, at $z=0$, given by
\begin{equation}
\label{eq:xPol}
\ve{E}_{\text{i}}=\ve{\hat{x}},
\quad
\ve{E}_{\text{r}}=S_{11}^{xx}\ve{\hat{x}} + S_{11}^{yx}\ve{\hat{y}},
\quad
\ve{E}_{\text{t}}=S_{21}^{xx}\ve{\hat{x}} + S_{21}^{yx}\ve{\hat{y}},
\end{equation}
where $S_{ab}^{uv}$, with $a, b = \{1,2\}$ and $u, v = \{x,y\}$, are the scattering parameters. The subscripts ``1'' and ``2'' refer to the regions $z<0$ and $z>0$, respectively. We may now define relations similar to~\eqref{eq:xPol} in the case of a $y$-polarized incident wave propagating in the positive $z$-direction and also $x$- and $y$-polarized incident waves propagating in the negative $z$-direction. Inserting the definition of all of these waves into~\eqref{eq:reducedSys}, allows us to define the matrices $\te{\Delta}$ and $\te{A}_v$, which are explicitly provided in Eqs.~\eqref{eq:DelAv} below and where we have used the following additional matrices:
\begin{subequations}
\label{eq:DelAv}
\begin{floatEq}
\begin{equation}
\label{eq:deltaMat}
\te{\Delta}=
\begin{pmatrix}
-\te{N}/\eta_1 + \te{N}\cdot\te{S}_{11}/\eta_1 + \te{N}\cdot\te{S}_{21}/\eta_2 & -\te{N}/\eta_2 +\te{N}\cdot\te{S}_{12}/\eta_1 + \te{N}\cdot\te{S}_{22}/\eta_2 \\
-\te{\text{J}}\cdot\te{N} -\te{\text{J}}\cdot\te{N}\cdot\te{S}_{11} +\te{\text{J}}\cdot\te{N}\cdot\te{S}_{21} &\te{\text{J}}\cdot\te{N} -\te{\text{J}}\cdot\te{N}\cdot\te{S}_{12}+\te{\text{J}}\cdot\te{N}\cdot\te{S}_{22}
\end{pmatrix},
\end{equation}
\begin{equation}
\label{eq:AvMat}
\te{A}_v=\frac{1}{2}
\begin{pmatrix}
\te{I} + \te{S}_{11}+ \te{S}_{21} &
\te{I} + \te{S}_{12}+ \te{S}_{22}
\\
\te{\text{J}}/\eta_1 -\te{\text{J}}\cdot\te{S}_{11}/\eta_1 +\te{\text{J}}\cdot\te{S}_{21}/\eta_2 &
-\te{\text{J}}/\eta_2 -\te{\text{J}}\cdot\te{S}_{12}/\eta_1 +\te{\text{J}}\cdot\te{S}_{22}/\eta_2
\end{pmatrix}.
\end{equation}
\end{floatEq}
\end{subequations}
\begin{equation}
\label{eq:defS}
\te{S}_{ab}=
\begin{pmatrix}
S_{ab}^{xx} & S_{ab}^{xy} \\
S_{ab}^{yx} & S_{ab}^{yy}
\end{pmatrix},\qquad
\te{N}=
\begin{pmatrix}
1 & 0 \\
0 & -1
\end{pmatrix}.
\end{equation}
The procedure to obtain the susceptibilities is thus as follows: 1) the scattering parameters for normal incidence of the metasurface are computed by full-wave simulations, 2) they are then used to define the matrices $\te{\Delta}$ and $\te{A}_v$ in~\eqref{eq:DelAv}, 3) the susceptibilities are finally found by matrix inversion of~\eqref{eq:reducedSys} along with~\eqref{eq:conv}.

A typical application of this procedure is that of relations~\eqref{eq:Xeexx} for a birefringent metasurface surrounded by vacuum, or in relations~\eqref{eq:BianiStoX} for a bianisotropic metasurface surrounded by two different media.

%
%
%

\begin{subequations}\label{eq:BianiStoX}
\begin{floatEq}
\begin{equation}
\chi_\text{ee}^{xx} = \chi_\text{ee}^{yy} = \frac{4j}{\epsilon\omega}\frac{(S_{11} - 1)(S_{22} - 1)-S_{21}^2}{\eta_1\left[S_{11}(S_{22}-1)+S_{22}-(1+S_{21})^2\right]+\eta_2\left[S_{11}(1+S_{22})-S_{22}-(1+S_{21})^2\right]},
\end{equation}
\begin{equation}
\chi_\text{mm}^{xx} = \chi_\text{mm}^{yy} = \frac{4j}{\mu\omega} \frac{\eta_1\eta_2(1+S_{11}+S_{22}+S_{11}S_{22}-S_{21}^2)}{\eta_1\left[S_{11}(S_{22}-1)+S_{22}-(1+S_{21})^2\right]+\eta_2\left[S_{11}(1+S_{22})-S_{22}-(1+S_{21})^2\right]},
\end{equation}
\begin{equation}
\chi_\text{em}^{xy} = -\chi_\text{em}^{yx}=\chi_\text{me}^{xy} = -\chi_\text{me}^{yx} = \frac{2j}{k} \frac{\eta_1\left[S_{22}+S_{11}(S_{22}-1)-(1+S_{21})^2\right]+\eta_2\left[S_{22}-S_{11}(1+S_{22})+(1+S_{21})^2\right]}{\eta_1\left[S_{11}(S_{22}-1)+S_{22}-(1+S_{21})^2\right]+\eta_2\left[S_{11}(1+S_{22})-S_{22}-(1+S_{21})^2\right]}.
\end{equation}
\end{floatEq}
\end{subequations}

\bibliographystyle{myIEEEtran}
\bibliography{NewLib}

\end{document}